\documentclass[a4paper,onecolumn,10pt,accepted=2021-08-24]{quantumarticle}
\pdfoutput=1

\usepackage{geometry}
\usepackage[english]{babel}
\usepackage[T1]{fontenc}
\usepackage{amsmath}
\usepackage{hyperref}
\usepackage{amsfonts}
\usepackage{bbm}
\usepackage{caption}
\usepackage{mathtools}
\usepackage{xspace}
\usepackage[numbers,sort&compress]{natbib}

\DeclareMathOperator*{\E}{\mathbb{E}}
\newcommand{\ket}[1]{|#1\rangle}
\newcommand{\bra}[1]{\langle #1|}
\newcommand{\braket}[1]{\langle #1 \rangle}
\newcommand{\braIket}[2]{\langle #1 | #2 \rangle}

\newcommand{\NN}{\mathbb{N}}
\newcommand{\QAOA}{\textsc{qaoa}\xspace}

\begin{document}

\title{Instance Independence of Single Layer Quantum Approximate Optimization Algorithm on Mixed-Spin Models at Infinite Size}
\author{Jahan Claes}
\affiliation{QC Ware Corporation, Palo Alto, CA USA}
\affiliation{Department of Physics and Institute for Condensed Matter Theory, University of Illinois at Urbana-Champaign, Urbana, IL 61801, USA}
\email{jclaes2@illinois.edu}
\author{Wim van Dam}
\affiliation{QC Ware Corporation, Palo Alto, CA USA}
\affiliation{Department of Computer Science, University of California, Santa Barbara, CA USA}
\affiliation{Department of Physics, University of California, Santa Barbara, CA USA}
\begin{abstract}
This paper studies the application of the Quantum Approximate Optimization Algorithm (\QAOA) to spin-glass models with random multi-body couplings in the limit of a large number of spins. 
We show that for such mixed-spin models the performance of depth $1$ \QAOA is independent of the specific instance in the limit of infinite sized systems and we give an explicit formula for the expected performance. 
We also give explicit expressions for the higher moments of the expected energy, thereby proving that the expected performance of \QAOA concentrates.  
\end{abstract}

\date{}

\maketitle

\tableofcontents

\section{Summary}
 The Quantum Approximate Optimization Algorithm (\QAOA) is a variational quantum algorithm designed to give approximate solutions to unconstrained binary optimization problems \cite{farhi2014quantum}. 
 While \QAOA can be proven to give the optimal answer in the limit where the number of \QAOA layers $p$ goes to infinity, rigorous results on the performance of \QAOA with finite $p$ are difficult to obtain. 
 In a recent paper, Farhi et al.\ \cite{farhi2019quantum} studied the application of the \QAOA to the Sherrington-Kirkpatrick (SK) model, a spin-glass model with random all-to-all two-body couplings, in the limit of a large number of spins. 
 Their paper demonstrated that for fixed $p$, the performance of the \QAOA is independent of the specific instance of the SK model and can be predicted by explicit formulas. 
 The paper also showed that the approximation ratio of the \QAOA at $p=11$ outperforms a large class of classical optimization algorithms (although not the best classical algorithm \cite{montanari2021optimization}). 
 In the current paper, we generalize the result of Farhi et al.\ to mixed-spin SK models, which extends the two-body couplings of standard SK to random all-to-all $q$-body couplings. 
 We demonstrate that for $p=1$, the performance of the \QAOA is again independent of the specific instance, and we provide an explicit formula for the expected performance. 
 Our work provides a potential avenue to demonstrating the advantage of \QAOA over classical algorithms, as the best known classical algorithms for mixed-spin SK models have an approximation ratio that is bounded away from $1$ \cite{alaoui2020optimization,alaoui2020algorithmic}.
 
 \section{Preliminaries and Notation}
The Quantum Approximate Optimization Algorithm (\QAOA) \cite{farhi2014quantum} is a heuristic quantum algorithm for binary optimization. 
Given a cost function of $n$ binary variables (spins) $H(z_1,\dots,z_n)$, \QAOA seeks to produce a string $z:=(z_1,\dots,z_n)$ close to the minimum of $H$. 
Commonly we view $H$ as a Hamiltonian operator that is diagonal in the $Z$-basis. 
A depth-$p$ \QAOA circuit then consists of $p$ repetitions of alternatively applying the Hamiltonian $H$ and the mixing Hamiltonian $B=X_1+\cdots +X_n$ to a uniform superposition as initial state, which is the product of $+X$ single particle eigenstates. 
Explicitly, the depth-$p$ \QAOA state is given by
\begin{equation}
    \ket{\beta_1,\dots,\beta_p; \gamma_1,\dots,\gamma_p} := e^{-i\beta_p B}e^{-i\gamma_p H}\cdots e^{-i\beta_1 B}e^{-i\gamma_1 H}\cdot \frac{1}{\sqrt{2^n}}\sum_{z\in\{\pm 1\}^n}\ket{z}.
\end{equation}
A depth-$p$ \QAOA circuit is parameterized by the $2p$ angles $\{\gamma_i\},\{\beta_i\}$. For a given problem, these angles should be optimized so that measuring in the $Z$-basis gives strings that make $H$ as small as possible. In practice, this is typically done by minimizing the expectation value of the energy
\begin{equation}
    \braket{H}  := \bra{\beta_1,\dots,\beta_p;\gamma_1,\dots,\gamma_p}H\ket{\beta_1,\dots,\beta_p;\gamma_1,\dots,\gamma_p}.
\end{equation}
For some problems, this minimization may be done analytically on a classical computer \cite{farhi2014quantum,jiang2017near,wang2018quantum,ozaeta2020expectation}. Otherwise, the minimization can be performed by running the \QAOA on a quantum computer repeatedly for a fixed set of angles, estimating the expectation value, and updating the angles according to classical minimization algorithms \cite{farhi2014quantum,crooks2018performance,zhou2020quantum}. We note that minimizing the energy expectation value is only one possible definition of ``best'' angles; in general minimizing the expectation value and maximizing the probability of finding the optimal $z$ (or maximizing the probability of $H(z_1,\dots,z_n)$ falling below a certain threshold) do not coincide.

It was recently demonstrated in \cite{brandao2018fixed} that in local optimization problems with cost functions drawn from from realistic random distributions (e.g.\ \textsc{MaxCut} on random 3-regular graphs), the expectation value per spin is \emph{instance independent} as $n\rightarrow\infty$. That is, for fixed angles $\{\gamma_i\}$ and $\{\beta_i\}$ $\braket{H/n}$ is the same for all problem instances. This implies that the angles $\{\gamma_i\}$ and $\{\beta_i\}$ do not need to be optimized for every problem instance, but can be optimized once and reused for every problem drawn from the same distribution. 
The methods of \cite{brandao2018fixed} can also be used to derive \emph{concentration of measure} results for local optimization problems. (While \cite{brandao2018fixed} did not explicitly address concentration of measure, it can be easily derived from their methodology.) That is, in the limit as $n\rightarrow\infty$, the variance in the energy per spin goes to zero:
\begin{equation}
\braket{(H/n)^2}-\braket{(H/n)}^2\rightarrow 0
\end{equation}
Concentration of measure means that for large $n$, every measurement of a \QAOA state in the $Z$-basis gives strings with the same energy per spin. In total, instance independence implies the \QAOA angles do not need to be optimized from instance to instance in order to minimize the expectation value, and concentration of measure implies that expectation value is the correct measure of the ``best" angles.

While instance independence (and concentration of measure) were initially derived for local cost functions, similar results have also been derived for the Sherrington-Kirkpatrick (SK) model \cite{sherrington1975solvable}, a physics-inspired optimization problem with cost function
\begin{equation}
    H = \sum_{1\leq j<k\leq n} J_{jk}Z_jZ_k
\end{equation}
with the $J_{jk}$ are independently drawn from a Gaussian distribution with mean $0$. The SK model is the ``most nonlocal" two-body cost function, and serves as a model for realistic nonlocal two-body optimization problems. In the limit $n\rightarrow\infty$, the ground state energy per spin is known to be independent of the instance and can be computed explicitly \cite{parisi1980sequence,talagrand2006parisi}. Recently, Montanari derived a classical algorithm that produces strings $z$ with energy within $(1-\epsilon)$ times the optimum; this method has a complexity of $C(\epsilon)n^2$, where $C(\epsilon)$ is a polynomial in $1/\epsilon$ \cite{montanari2021optimization}.

Ref.~\cite{farhi2019quantum} proved both instance independence and concentration of measure for depth-$p$ \QAOA applied to the SK model. In addition, \cite{farhi2019quantum} provided an explicit formula for $\braket{H/n}$ in the limit $n\rightarrow\infty$ for $p=1$, and provided a computer algorithm to generate $\braket{H/n}$ for any fixed depth $p>1$. Therefore, the \QAOA angles for the SK model can be chosen on a classical computer, and there are fixed performance guarantees in the limit $n\rightarrow\infty$. Ref.~\cite{farhi2019quantum} demonstrated that at $p=11$, \QAOA outperforms semidefinite programming for the SK model, but could not show that the \QAOA matches the performance of the Montanari algorithm.

In this work, we study a generalization of the SK model, the mixed-spin SK model, that allows for polynomials of degree $d$ in the binary variables instead of only degree-$2$ terms \cite{panchenko2012sherrington,panchenko2013sherrington}. This can serve as a model for nonlocal optimization problems with higher-order terms. The mixed-spin SK model is also known to have a ground state energy per spin that is independent of the instance and can be computed explicitly \cite{panchenko2012sherrington, auffinger2017parisi,panchenko2013sherrington,panchenko2014parisi}. For a mixed-spin SK model with degree $d=3$, the generalization of the Montanari algorithm \cite{alaoui2020optimization} approaches a fixed approximation ratio of $\sim(0.9843\pm 0.0003)$ times the optimum value, rather than the optimal value \cite{alaoui2020algorithmic}. Thus, the mixed-spin SK model is a potential avenue for establishing the advantage of the \QAOA over classical algorithms.

In this paper, we generalize the work of \cite{farhi2019quantum} to prove instance independence and concentration of measure for $p=1$ \QAOA applied to mixed-spin SK models. As part of our work, we derive an explicit formula for $\braket{H/n}$ in the limit $n\rightarrow\infty$, implying that the \QAOA angles for the mixed-spin SK models can be chosen on a classical computer. Our work can likely be generalized to depth $p\geq 1$ using the same methods as \cite{farhi2019quantum}.

\section{Current Results and Prior Work}
\subsection{Our Results on Mixed-Spin Sherrington-Kirkpatrick Model}
The mixed-spin SK model (often called the mixed $p$-spin model, although we will not use this terminology to avoid confusion with the $p$ of \QAOA) is given by \cite{panchenko2012sherrington,panchenko2013sherrington}
\begin{align}
\begin{split}
H & = \sum_{j=1}^n J_j Z_j+\frac{1}{\sqrt{n}}\sum_{1\leq j<k \leq n}J_{jk}Z_jZ_k+\frac{1}{n}\sum_{1\leq j<k<\ell\leq n}J_{jk\ell}Z_jZ_kZ_\ell +\cdots +\frac{1}{n^{\frac{d-1}{2}}}\sum_{1\leq j_1<\cdots j_d \leq n}J_{j_1\cdots j_d} Z_{j_1}\cdots Z_{j_d} \\
& = \sum_{q=1}^d n^{\frac{1-q}{2}}\sum_{\substack{S\subseteq\{1,\dots,n\}\\|S|=q}} J_SZ_S \label{eq:MixedSKModelDefn}
\end{split}
\end{align}
where in the last line we denote the product $\prod_{i\in S}Z_i$ by $Z_S$. We assume each $J_S$ is sampled independently from a Gaussian distribution $N(0,\sigma^2_{|S|})$ that only depends on $|S|$, and we will let $\sigma_q$ be the standard deviation of the coupling constants $J_S$ with $|S|=q$, for $q=1,\dots,d$.

The ground state of the mixed-spin SK Hamiltonian is known to have a fixed energy per spin in the limit $n\rightarrow\infty$ \cite{panchenko2014parisi}. In fact we can also allow arbitrarily high orders in the mixed-spin SK model ($d=\infty$), provided the variances decrease quickly enough to make $\sum_q 2^q\sigma_q^2$ finite, and the ground state model will still have a fixed energy per spin as $n\rightarrow\infty$ \cite{panchenko2014parisi}. However, for simplicity we consider some finite bound $d$ on the degree.

Our main result is as follows. 
Define the $n$-spin model by Eq.~\ref{eq:MixedSKModelDefn} with $J_{S} \sim N(0,\sigma^2_{|S|})$. 
Then, using depth $1$ \QAOA with angles $\beta,\gamma$, the expectation of the energy per spin equals in the large $n$ limit is given by 
\begin{align}
    \lim_{n\rightarrow\infty}\E_{J\sim N}\left[\braket{H/n}\right]&=2\gamma\sum_{q=1}^d\sigma_q^2\sum_{\substack{a\text{ odd}\\a\leq q}}\frac{(-1)^{\frac{a-1}{2}}}{a!(q-a)!}\exp\left(-2\gamma^2a\sum_{q'=1}^d  \frac{\sigma_{q'}^2}{(q'-1)!}\right)\sin^a(2\beta)\cos^{q-a}(2\beta),\label{eq:firstMomentResult}.
\end{align}
If we define the variables $c_q:=\sigma_q/\sqrt{q!}$ and the polynomial $\xi(x):=\sum_{q=1}^d c_q^2 x^q$, as is common in discussions of the mixed-spin model (see Section  \ref{sec:PriorResults}), we can also write this as
\begin{align}
    \lim_{n\rightarrow\infty}\E_{J\sim N}\left[\braket{H/n}\right]=
    2\gamma\Im\left[\xi 
    \Big(
       \cos(2\beta) + 
    i \sin(2\beta) \exp\left({-2\gamma^2 \xi'(1)}\right)
    \Big)\right].\label{eq:firstMomentResultMontanariForm}
\end{align}
Furthermore, the second moment of the energy per spin equals
\begin{align}
    \lim_{n\rightarrow\infty}\E_{J\sim N}\left[\braket{(H/n)^2}\right]&=\lim_{n\rightarrow\infty}\E_{J\sim N}\left[\braket{H/n}\right]^2.\label{eq:SecondMomentResult}
\end{align}

Our second result, Eq.~\ref{eq:SecondMomentResult}, allows us to prove that $p=1$ \QAOA applied to the mixed-spin SK model has both concentration of measure and instance independence. 
To see this, we note that we may write
\begin{align}
    \E_{J\sim N}\left[\braket{(H/n)^2}\right]-\E_{J\sim N}\left[\braket{H/n}\right]^2& =\Big(\E_{J\sim N}\left[\braket{(H/n)^2}-\braket{H/n}^2\right]\Big)+\Big(\E_{J\sim N}\left[\braket{H/n}^2\right]-\E_{J\sim N}\left[\braket{H/n}\right]^2\Big)\label{eq:TotalVariance}
\end{align}
When applying \QAOA to mixed-spin SK models, the measured value of $(H/n)$ varies for two reason. 
First, it varies because the bonds $J_S$ vary from instance to instance (the $\E_J\left[\cdot\right]$ expectation). 
Second, it varies because the \QAOA state $\ket{\gamma,\beta}$ is not an eigenstate of the $Z$-operators, so that the measurement outcomes have randomness even for fixed $J_S$ (the $\braket{\cdot}$ expectation). 
The left hand side of Eq.~\ref{eq:TotalVariance} represents the total variance in $(H/n)$ due to both sources of randomness. 
The right hand side demonstrates that the total variance can be decomposed into two terms. 
The first term is the average over $J_S$ of the variance due only to the measurement randomness. 
The second term is the variance in the expected value $\braket{H/n}$ due to the randomness in the bonds $J_S$. 
Since both of these terms are non-negative, they must both tend to zero as $n\rightarrow\infty$ as well.

The fact that the first variance approaches zero gives us concentration of measure: it says that for typical couplings $J_S$ the variance in the measurement outcomes vanishes, and thus we always measure a string with energy per spin equal to the expectation value (note that the term inside the $\E\left[\cdot\right]$ is always positive, so that for the average over $J_S$ to go to zero the magnitude of the typical value must also go to zero). 
The second term approaching zero clearly gives instance independence of the expectation value, since it shows that the variance in the expectation value due to different couplings vanishes.

Finally, we note that the methods we use also suffice to derive a formula for all higher moments of the energy per spin
\begin{align}
    \lim_{n\rightarrow\infty}\E_{J\sim N}\left[\braket{(H/n)^m}\right]&=\lim_{n\rightarrow\infty}\E_{J\sim N}\left[\braket{H/n}\right]^m,\label{eq:mthMomentResult}
\end{align}
although unlike the $m=2$ result, the $m>2$ result does not have any obvious implication for the performance of the \QAOA.

\subsection{Numerical Results and Relation to Prior Results}
\label{sec:PriorResults}

\begin{figure}
    \centering
    \includegraphics[width=\columnwidth]{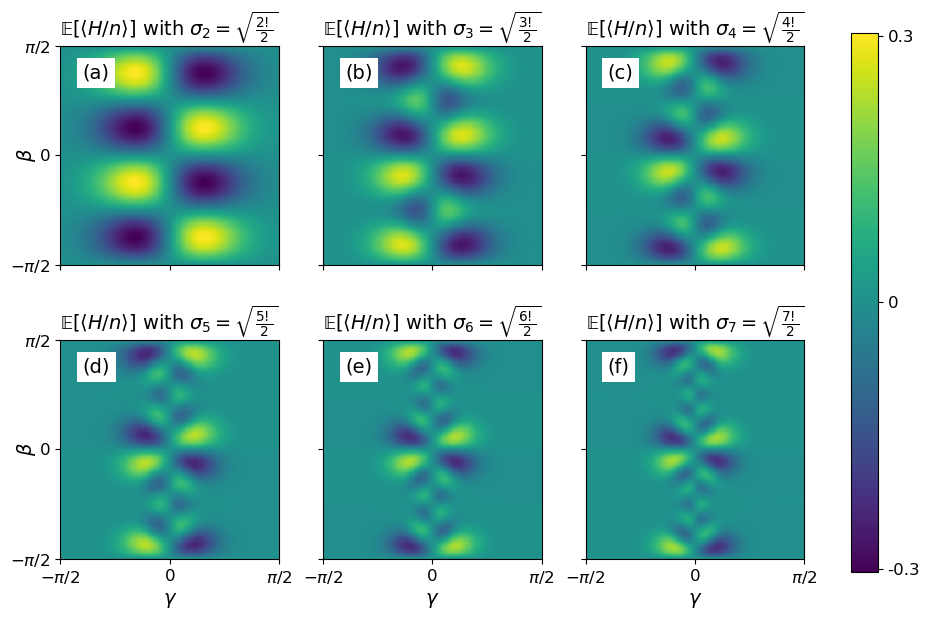}
    \caption{The energy landscapes for the pure $d$-spin models with $\sigma_q=\delta_{d,q}\sqrt{d!/2}$, in the limit as $n\rightarrow\infty$.}
    \label{fig:ExpectationValues}
\end{figure}

\begin{figure}
    \centering
    \includegraphics[width=\columnwidth]{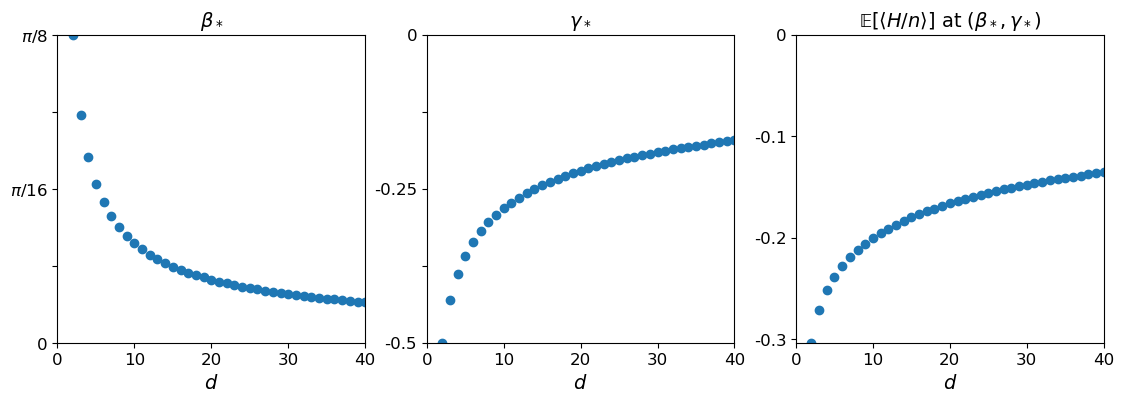}
    \caption{The optimal angles $(\beta_*,\gamma_*)$ for the pure $d$-spin models with $\sigma_q=\delta_{d,q}\sqrt{d!/2}$ that minimize the energy per spin $\mathbb{E}\left[\braket{H/n}\right]$ in the limit $n\rightarrow\infty$.}
    \label{fig:optimalAngles}
\end{figure}

As an example of using our Eqs.\ \ref{eq:firstMomentResult} and \ref{eq:firstMomentResultMontanariForm}, we can compute the optimal angles for pure $d$-spin models given by $\sigma_q=\delta_{d,q}\sqrt{d!/2}$. 
In this case, our central equation reduces to 
\begin{align}
\begin{split}
    \lim_{n\rightarrow\infty}\E_{J\sim N}\left[\braket{H/n}\right]&=\gamma d!\sum_{\substack{a\text{ odd}\\a\leq d}}\frac{(-1)^{\frac{a-1}{2}}}{a!(d-a)!}\exp\left(-2\gamma^2ad\right)\sin^a(2\beta)\cos^{d-a}(2\beta)\\
    &=\gamma\Im\left[\Big(\cos(2\beta)+i\sin(2\beta)\exp\left({-2\gamma^2d}\right)\Big)^d\right].
\end{split}
\end{align}
Examples of this energy landscape for small $d$ are plotted in Fig.~\ref{fig:ExpectationValues}, and numerically optimized angles and corresponding energy per spin are plotted in  Fig.~\ref{fig:optimalAngles}.

\begin{figure}
    \centering
    \includegraphics[width=\columnwidth]{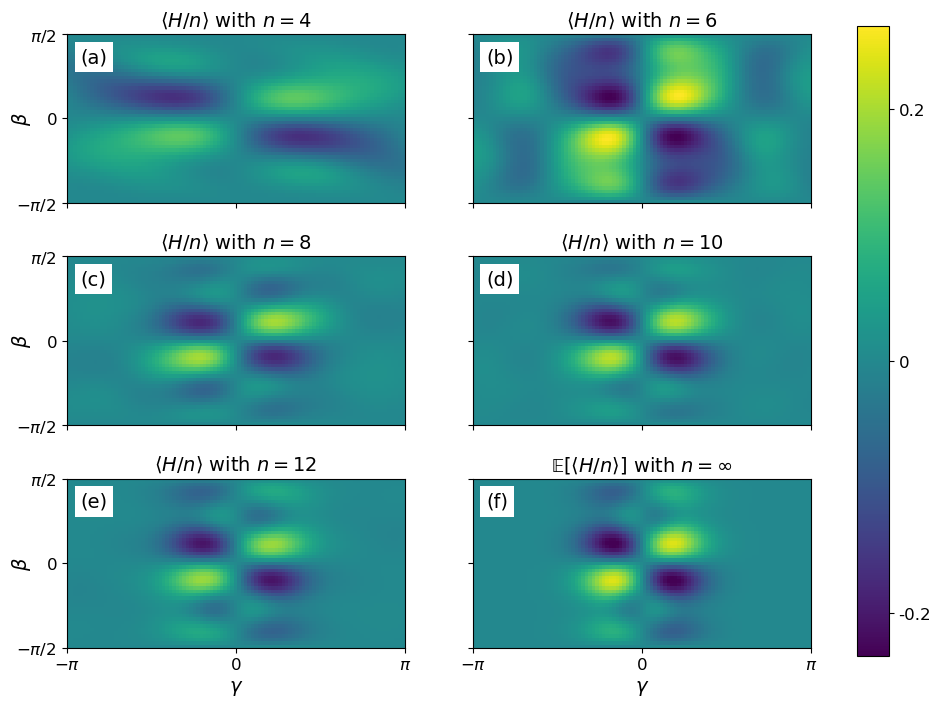}
    \caption{(a-e) The energy landscape of random instances of a mixed-spin optimization problem with $(\sigma_1,\sigma_2,\sigma_3)=(1/3,1/2,1)$, at varying problem sizes $n$. 
    (f) The predicted energy landscape in the limit $n\rightarrow\infty$. 
    We see the energy landscape for random instances rapidly converges to the $n\rightarrow\infty$ landscape, demonstrating instance independence in this limit.}
    \label{fig:ExpectationValuesConvergence}
\end{figure}

As another example application of our formula, we explore the instance independence at finite $n$ to demonstrate convergence as $n\rightarrow\infty$. In Fig.~\ref{fig:ExpectationValuesConvergence}a-e, we plot the expected energy per spin $\braket{H/n}$ for five randomly generated mixed-spin optimization problems generated from a distribution with $(\sigma_1,\sigma_2,\sigma_3)=(1/3,1/2,1)$ (and all higher-order terms zero). We see that as we increase $n$, the energy landscape of any given problem instance quickly approaches the $n\rightarrow\infty$ landscape (Fig.~\ref{fig:ExpectationValuesConvergence}f), as predicted.

To put Eq.~\ref{eq:firstMomentResult} in perspective we explicitly calculate what it implies for two basic cases $q=2$ and $q=3$ and compare it with the literature on mixed-spin models. 

The standard Sherrington-Kirkpatrick model has $q=2$, $\sigma_2=1$, and $\sigma_q=0$ for all other $q\neq 2$. 
In this case our Eq.~\ref{eq:firstMomentResult} tells us that 
\begin{align}
    \E_{J\sim N}\left[\braket{H_{SK}/n }\right] & =2\gamma e^{-2\gamma^2}\sin(2\beta)\cos(2\beta)=\gamma e^{-2\gamma^2}\sin(4\beta),
\end{align}
which agrees with previous work in \cite{farhi2019quantum} and \cite{ozaeta2020expectation}.
This expectation is minimized by the parameters $\beta_*=\pm\pi/8 \approx \pm 0.392699$ and $\gamma_* = \mp 0.5$ for which $\E\left[\braket{H_{SK}/n }\right] = -1/\sqrt{4e}\approx -0.303265$. 
This is the same result we found numerically for $d=2$ in Fig.~\ref{fig:optimalAngles}.


For the more general case $q\geq 2$ we will briefly review the notation used by the articles \cite{alaoui2020algorithmic,alaoui2020optimization,panchenko2014parisi,auffinger2017parisi}.
In the current paper the mixed-spin model is described by the Hamiltonian
\begin{align}
H & = \sum_{q=1}^d\frac{1}{\sqrt{n^{q-1}}}\sum_{1\leq j_1 <\cdots < j_q\leq n} {J_{j_1\cdots j_q}Z_{j_1}\cdots Z_{j_q}},
\end{align}
with $J_{j_1\cdots j_q}\sim N(0,\sigma_q^2)$ and for a fixed $q$ the summation has $\binom{n}{q}$ terms.
In \cite{alaoui2020algorithmic} and elsewhere a different notation is used where the Hamiltonian is defined by 
\begin{align}
H' & = 
\sum_{q=1}^d\frac{c_q}{\sqrt{n^{q-1}}}\sum_{1\leq j_1,\dots,j_q \leq n} {J_{j_1\cdots j_q}Z_{j_1}\cdots Z_{j_q}},
\end{align}
with $J_{j_1\cdots j_q}\sim N(0,1)$. 
Crucially, now the indices $j_1,\dots,j_q$ can have repeated values and permutations are treated distinctly, hence for each $q$ the summation involves $n^q$ terms. 

In the limit of large $n$ one can show that the $q$-plets 
$(j_1,\dots j_q)$ with repeated values can be ignored as their relative contribution decreases as a function $n$. 
For distinct indices there are $q!$ permutations to consider in $H'$, hence we have a sum of $q!$ standard normal distributions: 
$J'_{j_1 \dots j_q} + \cdots + J'_{j_q\cdots j_1}$, which is identical to a single normal distribution with variance  $q!$.  
As a result we re-express $H'$ as 
\begin{align}
H' & = 
\sum_{q=1}^d\frac{c_q \sqrt{q!}}{\sqrt{n^{q-1}}}\sum_{1\leq j_1<\dots<j_q \leq n} {J_{j_1\cdots j_q}Z_{j_1}\cdots Z_{j_q}},
\end{align}
with $J\sim N(0,1)$. 
Thus when $\sigma_q = c_q \sqrt{q!}$ we have that $H$ and $H'$ describe the same model.
It is common to capture the different $c_q$ coefficients by the \emph{mixture} function $\xi(x) = \sum_q c_q^2x^q$, such that the standard SK model has $\xi(x) = x^2/2$ (i.e.\ $c_2=1/\sqrt{2}$ and hence $\sigma_2=1$).

For the $3$-body model $H_3$, \cite{alaoui2020optimization} analyzes the mixture $\xi(x) = c_3^2x^3 = x^3/2$, such that $c_3 = 1/\sqrt{2}$.
It was shown that the expected ground state energy-per-spin of this model equals $-0.8132 \pm 0.0001$ 
In the notation of the current article this model is equivalent to $\sigma_3=\sqrt{3}$ and with this our Eq.~\ref{eq:firstMomentResult} gives the expectation 
\begin{align}
    \E_{J\sim N}\left[\braket{H_3/n }\right] & 
    = 3\gamma e^{-3\gamma^2}\sin(2\beta)\cos^2(2\beta)-\gamma e^{-9\gamma^2}\sin^3(2\beta).
\end{align}
This expectation is minimized to $\E_J[\braket{H_3/n }]\approx -0.270638$ by the angles $(\beta_*,\gamma_*)\approx (\pm 0.290003,\mp 0.430091)$, which are the solutions to 
\begin{align}
\beta_* & = -\frac{1}{4}\arccos\left(1-\frac{1}{9\gamma_*^2}\right) \\
\exp(-6\gamma_*^2)  & = 18\gamma_*^2-3 \\
\E_J[\braket{H_3/n }] & = \sqrt{4\gamma_*^2-\frac{2}{3}}.
\end{align}
This is the same result we found numerically for $d=3$ in Fig.~\ref{fig:optimalAngles}. 
We thus see how depth $p=1$ \QAOA can approximate the ground state energy by a factor of $0.332806$. This $0.33$ approximation factor had been reported earlier by Zhou et al.\cite{zhouQIP2020}

\section{Derivation of Main Result}

The proof in this section follows to large extent the framework of the earlier result by Farhi et al.~\cite{farhi2019quantum}, which relied on manipulating the moment generating function $\E_J\left[\braket{e^{i\lambda H/n}}\right]$ to extract expressions for the first and second moments of $(H/n)$. 
We use their method of simplifying the moment generating function, and their reorganization of the sum over $z$-strings into a sum over sketches (see Section~\ref{sec:sketches}). 
We extend their proof technique by generalizing their form of the moment generating function to higher-spin models and demonstrating that it can still be written as a sum over sketches, developing careful power-counting methods to allow us to extract the relevant terms in the $n\rightarrow\infty$ limit, and deriving identities that allow us to explicitly evaluate the relevant sums.

In our proof, we will use the following conventions:
\begin{itemize}
    \item A $Z$-basis state $\ket{z}$ is specified by a string $z=(z_1,z_2,\dots, z_n)$ of $\pm 1$s. 
    We will use the shorthand $z\in\{\pm 1\}^n$ for this. 
    \item The XOR of two bitstrings $z$ and $z'$ is given by the componentwise product  $zz':=(z_1z'_1,z_2z'_2,\dots,z_nz_n')\in\{\pm 1\}^n$.
    \item For a set $S\subseteq \{1,\dots,n\}$, we denote the product of the bits in $S$ as $\prod_{i\in S}z_i=:z_S$; with this convention we thus have $Z_S\ket{z}=z_S\ket{z}$.
    \item The uniform superposition over all strings $z$ is denoted by 
    \begin{align}
    \ket{\Sigma} & := \frac{1}{\sqrt{2^n}}\sum_{z\in\{\pm 1\}^n}\ket{z}. 
    \end{align}
\end{itemize}
This is in contrast to the usual convention in quantum information, in which a $Z$-basis state is specified by a string $z$ of $0$s and $1$s and the XOR of two strings is given by componentwise addition modulo 2. We choose our notation to be consistent with \cite{farhi2019quantum} and to simplify certain expressions in our derivation.

\subsection{Moment Generating Function}

Following \cite{farhi2019quantum}, to evaluate $\E_J[\braket{H/n}]$ and $\E_J[\braket{(H/n)^2}]$ we use the moment-generating function $\E_J[\braket{e^{i\lambda H/n}} ]$.
From the moment-generating function, we can find the moments via
\begin{align}
    \E_{J\sim N}\left[\braket{(H/n)^m}\right]
    &=\left.(-i\partial_\lambda)^m     \E_{J\sim N}\left[\braket{e^{i\lambda H/n}}\right]\right|_{\lambda=0}\label{eq:momentDefn}
\end{align}

We can simplify the expectation inside the moment-generating function by inserting three complete sets of states:
\begin{align}
\begin{split}
    \braket{e^{i\lambda H/n}} 
    &= \bra{\Sigma}e^{i\gamma H}e^{i\beta B} e^{i\lambda H/n}e^{-i\beta B}e^{-i\gamma H}\ket{\Sigma}\\
    &= \sum_{z,z',z''\in\{\pm 1\}^n}\braIket{\Sigma}{z}\bra{z}e^{i\gamma H}e^{i\beta B} \ket{z''}\bra{z''}e^{i\lambda H/n}e^{-i\beta B}e^{-i\gamma H}\ket{z'}\braIket{z'}{\Sigma}\\
    &= \frac{1}{2^n}\sum_{z,z',z''\in\{\pm 1\}^n} \exp{\left(i\gamma \left[H(z)-H(z')\right]+i\frac{\lambda H(z'')}{n}\right)} \bra{z}e^{i\beta B}\ket{z''}\bra{z''}e^{-i\beta B}\ket{z'}\\
    &= \frac{1}{2^n}\sum_{z,z',z''\in\{\pm 1\}^n}\exp{\left(i\sum_{q=1}^d n^{\frac{1-q}{2}}\sum_{|S|=q}J_S\left[\gamma (z_S-z'_S)+\frac{\lambda z''_S}{n}\right]\right)}\bra{z}e^{i\beta B}\ket{z''}\bra{z''}e^{-i\beta B}\ket{z'}\\
    &= \frac{1}{2^n}\sum_{z,z',z''\in\{\pm 1\}^n}\exp{\left(i\sum_{q=1}^d n^{\frac{1-q}{2}}\sum_{|S|=q}z''_S J_S\left[\gamma (z_S-z'_S)+\frac{\lambda}{n}\right]\right)}\bra{z}e^{i\beta B}\ket{(+1)^n}\bra{(+1)^n}e^{-i\beta B}\ket{z'},
\end{split}
\end{align}
where in the last step we used that $\bra{z}e^{i\beta B}\ket{z'}=\bra{zz'}e^{i\beta B}\ket{(+1)^n}$ and made the replacements $z\mapsto zz''$ and $z'\mapsto z'z''$. 

\subsection{Expectation when Couplings are Normal Distributed Variables}
\label{sec:sketches}
We will now treat the $J_S$ couplings as a random variable and consider the expectation $\E_J$ of the energy. 
We assume that the distribution is symmetric with respect to $J \leftrightarrow -J$, such that we can replace $z''_S J_S$ by $J_S$ to get  
\begin{align}
\begin{split}
    \E_{J}\left[\braket{e^{i\lambda H/n}}\right]
    &= \frac{1}{2^n}\sum_{z,z',z''\in\{\pm 1\}^n}\E_{J\sim N}\left[\exp{\left(i\sum_{q=1}^d n^{\frac{1-q}{2}}\sum_{|S|=q}z''_S J_S\left[\gamma (z_S-z'_S)+\frac{\lambda}{n}\right]\right)}\right]\bra{z}e^{i\beta B}\ket{(+1)^n}\bra{(+1)^n}e^{-i\beta B}\ket{z'}\\
    &= \sum_{z,z'\in\{\pm 1\}^n}\E_{J\sim N}\left[\exp{\left(i\sum_{q=1}^d n^{\frac{1-q}{2}}\sum_{|S|=q}J_S\left[\gamma (z_S-z'_S)+\frac{\lambda}{n}\right]\right)}\right]\bra{z}e^{i\beta B}\ket{(+1)^n}\bra{(+1)^n}e^{-i\beta B}\ket{z'}.
\end{split}
\end{align}
When we further assume that the $J_S$ variables are independent between different $S$ we can continue by 
\begin{align}
    \E_{J}\left[\braket{e^{i\lambda H/n}}\right]
    &= \sum_{z,z'\in\{\pm 1\}^n}\prod_{q=1}^d \prod_{|S|=q}\E_{J_S}\left[\exp{\left(i\cdot n^{\frac{1-q}{2}}J_S\left[\gamma (z_S-z'_S)+\frac{\lambda}{n}\right]\right)}\right]\bra{z}e^{i\beta B}\ket{(+1)^n}\bra{(+1)^n}e^{-i\beta B}\ket{z'}.
\end{align}

Next we assume that the $J_S$ are normally distributed with a standard deviation that is the same for all sets $S$ of the same size, that is $J_S\sim N(0,\sigma^2_{|S|})$. 
We note that taking the expectation value of a Gaussian random variable $J$ with standard deviation $\sigma$ gives
\begin{align}
\E_{J\sim N(0,\sigma^2)}\left[ e^{icJ}\right]=\frac{1}{\sigma\sqrt{2\pi}}\int e^{-J^2/2\sigma^2+icJ}=e^{-c^2\sigma^2/2}
\end{align}
so that our overall expression becomes
\begin{align}
    \E_{J\sim N}\left[\braket{e^{i\lambda H/n}}\right] &=\smashoperator[r]{\sum_{z,z'\in \{\pm 1\}^n}}\exp{\left(-\sum_{q=1}^d \frac{\sigma_{q}^2}{2n^{q-1}}\sum_{|S|=q}\left[\gamma^2 (z_S-z'_S)^2+\frac{2\gamma\lambda(z_S-z'_S)}{n}+\frac{\lambda^2}{n^2}\right]\right)}\bra{z}e^{i\beta B}\ket{(+1)^n}\bra{(+1)^n}e^{-i\beta B}\ket{z'}.
\end{align}
To do the sum over $z,z'$, we claim that the summand does not depend on all $2n$ spin values of $z$ and $z'$. 
Instead, it is only a function of the four integer values $(n_{++},n_{+-},n_{-+},n_{--})$, where $n_{ss'}$ is defined to be the number of positions $k\in \{1,\dots,n\}$ with $(z_k,z'_k)=(s,s')$.
Note that only three of these variables are actually independent, as we always have $n_{++}+n_{+-}+n_{-+}+n_{--}=n$.
As these numbers summarize the crucial information of the strings, we will refer to $(n_{++},n_{+-},n_{-+},n_{--})$ as the \emph{sketch} of $(z,z')$. 
Writing the summand in terms of the sketch rather than $(z,z')$ was introduced in \cite{farhi2019quantum}; here we establish that we can still write the summand in terms of the sketch for the mixed-spin SK model.
To start, it is straightforward to verify that
\begin{align}
    \bra{z}e^{i\beta B}\ket{(+1)^n}\bra{(+1)^n}e^{-i\beta B}\ket{z'} &=\smashoperator[r]{\prod_{ss'\in\{\pm\}^2}} Q_{ss'}^{n_{ss'}},
    \text{~with~}
\begin{cases}    
Q_{++} :=\cos^2(\beta)\\ 
Q_{--}:=\sin^2(\beta) \\ 
Q_{-+}:=i\sin(\beta)\cos(\beta)\\ 
Q_{+-}:=-i\sin(\beta)\cos(\beta).\label{eq:QDefinition}
\end{cases}
\end{align}
We can also write explicit combinatorial formulas for the sums in the exponential:
\begin{align}
\begin{split}
\sum_{\substack{S\subseteq\{1,\dots,n\}\\|S|=q}}(z_S-z_S')&=\sum_{k=0}^q(-1)^{q-k}\left[\binom{n_{++}+n_{+-}}{k}\binom{n_{-+}+n_{--}}{q-k}-\binom{n_{++}+n_{-+}}{k}\binom{n_{+-}+n_{--}}{q-k}\right]\\
&=\sum_{k=0}^q(-1)^{q-k}\left[\binom{\frac{n+(n_{+-}-n_{-+})+(n_{++}-n_{--})}{2}}{k}\binom{\frac{n-(n_{+-}-n_{-+})-(n_{++}-n_{--})}{2}}{q-k}\right.\\&\qquad\qquad\qquad\qquad-\left.\binom{\frac{n-(n_{+-}-n_{-+})+(n_{++}-n_{--})}{2}}{k}\binom{\frac{n+(n_{+-}-n_{-+})-(n_{++}-n_{--})}{2}}{q-k}\right]\\
&=: f_q(n_{++}, n_{+-}, n_{-+}, n_{--})\label{eq:fqDefinition}
\end{split}\\
\begin{split}
\sum_{\substack{S\subseteq\{1,\dots,n\}\\|S|=q}}(z_S-z_S')^2 &= 4\sum_{k\text{ odd}} \binom{n_{+-}+n_{-+}}{k}\binom{n_{++}+n_{--}}{q-k}\\
&=4\sum_{k\text{ odd}} \binom{n_{+-}+n_{-+}}{k}\binom{n-(n_{+-}+n_{-+})}{q-k}\\
&=: g_q(n_{++}, n_{+-}, n_{-+}, n_{--}).\label{eq:gqDefinition}
\end{split}
\end{align}
Therefore, the summand indeed depends only on $(n_{++}, n_{+-}, n_{-+}, n_{--})$ and the number of ways to assign the $n$ positions into four groups of these sizes is the multinomial  $n!/(n_{++}! n_{+-}! n_{-+}! n_{--}!)$. 
To condense our notation we will use $\{n_*\}$ to denote the set of sketches $n_* = (n_{++}, n_{+-}, n_{-+}, n_{--})$, allowing us to use the shorthand
\begin{align}
\sum_{\{n_*\}}\binom{n}{n_*} F(n_*) & := 
\sum_{\substack{(n_{++}, n_{+-}, n_{-+}, n_{--})\in\NN^4 \\ n_{++}+n_{+-}+n_{-+}+n_{--} = n}}\binom{n}{n_{++}, n_{+-}, n_{-+}, n_{--}}F(n_{++}, n_{+-}, n_{-+}, n_{--}).
\end{align}
Note that this summation has $\binom{n+3}{3}$ terms.
We thus have
\begin{align}
    \E_{J\sim N}\left[\braket{e^{i\lambda H/n}}\right] &= \sum_{\{n_*\}}\binom{n}{n_*}\exp\left({-\sum_{q=1}^d \left[\gamma^2 g_q(n_*)+2\gamma\lambda \frac{f_q(n_*)}{n}+\lambda^2\frac{\left(\begin{smallmatrix}n\\q\end{smallmatrix}\right)}{n^2}\right]\frac{\sigma_{q}^2}{2n^{q-1}}}\right) \smashoperator[r]{\prod_{ss'\in\{\pm\}^2}} Q_{ss'}^{n_{ss'}}.\label{eq:momentGeneratingFinalForm}
\end{align}
Eq.~\ref{eq:momentGeneratingFinalForm} is the form of the moment-generating function we will use to evaluate $\E_J[\braket{H/n}]$ and $\E_J[\braket{(H/n)^2}]$. 

\subsection{General Form of Moments}

Using Eq.~\ref{eq:momentDefn} combined with the form of the moment-generating function given in Eq.~\ref{eq:momentGeneratingFinalForm}, we can write the first moment as:
\begin{align}
    \E_{J\sim N}\left[\braket{H/n}\right] & = i\gamma\sum_{q=1}^d\frac{\sigma_q^2}{n^q}\sum_{\{n_*\}}\binom{n}{n_*}\exp\left({-\sum_{q'=1}^d \gamma^2 g_{q'}(n_*)\frac{\sigma_{q'}^2}{2n^{q'-1}}}\right) f_q(n_*)\smashoperator[r]{\prod_{ss'\in\{\pm\}^2}} Q_{ss'}^{n_{ss'}},
\end{align}
and the second moment as:
    \begin{align}
    \E_{J\sim N}\left[\braket{(H/n)^2}\right] =&
    \sum_{q=1}^d\binom{n}{q}\frac{\sigma_{q}^2}{n^{{q}+1}}-\gamma^2\sum_{q,q'=1}^d\frac{\sigma_q^2\sigma_{q'}^2}{n^{q+q'}}\sum_{\{n_*\}}\binom{n}{n_*}
    \exp\left({-\sum_{q''=1}^d \gamma^2 g_{q''}(n_*)\frac{\sigma_{q''}^2}{2n^{q''-1}}}\right)  f_q(n_*)f_{q'}(n_*)\smashoperator[r]{\prod_{ss'\in\{\pm\}^2}} Q_{ss'}^{n_{ss'}}.
\end{align}
The explicit expression for $f_q$ (Eq.~\ref{eq:fqDefinition}) shows that $f_q$ is a degree-$q$ polynomial in the variables $(n_{+-}-n_{-+})$, $(n_{++}-n_{--})$, and $n$, so we can expand $f_q$ as
\begin{equation}
    f_q(n_*)=\sum_{a+b+c\leq q} f_q^{abc}(n_{+-}-n_{-+})^a(n_{++}-n_{--})^bn^c\label{eq:fqExpansion}
\end{equation}
where the $f^{abc}_q$ are constants independent of $n$. 
In terms of this expansion, we have
\begin{align}
    \E_{J\sim N}\left[\braket{H/n}\right] &=i\gamma\sum_{q=1}^d\sigma_q^2\sum_{a+b+c\leq q}\frac{f_q^{abc}}{n^{q-c}}\sum_{\{n_*\}}\binom{n}{n_*}\exp\left(-\sum_{q'=1}^d  g_{q'}(n_*)\frac{\gamma^2\sigma_{q'}^2}{2n^{q'-1}}\right) (n_{+-}-n_{-+})^a(n_{++}-n_{--})^b\smashoperator[r]{\prod_{ss'\in\{\pm\}^2}} Q_{ss'}^{n_{ss'}}\label{eq:firstMomentForm}
\end{align}
    and 
\begin{align}
\begin{split}
    \E_{J\sim N}\left[\braket{(H/n)^2}\right] & = \sum_{q=1}^d\binom{n}{q}\frac{\sigma_{q}^2}{n^{q+1}} 
-\gamma^2\sum_{q,q'=1}^d\sigma_q^2\sigma_{q'}^2\sum_{\substack{a+b+c\leq q\\a'+b'+c'\leq q'}}\frac{f^{abc}_q}{n^{q-c}}\frac{f^{a'b'c'}_{q'}}{n^{q'-c'}}\sum_{\{n_*\}}\binom{n}{n_*}\exp\left(-\sum_{q''=1}^d  g_{q''}(n_*)\frac{\gamma^2\sigma_{q''}^2}{2n^{q''-1}}\right)\label{eq:secondMomentForm} \\& \qquad\qquad\qquad\qquad\qquad\qquad\qquad\qquad\qquad\qquad \times (n_{+-}-n_{-+})^{a+a'}(n_{++}-n_{--})^{b+b'}\smashoperator[r]{\prod_{ss'\in\{\pm\}^2}} Q_{ss'}^{n_{ss'}}.
\end{split} 
\end{align}
Ref.~\cite{farhi2019quantum} could explicitly evaluate these terms for the small values of $a$ and $b$ relevant for the two-body SK model, using concise expressions for $f_2$ and $g_2$. 
However, to get explicit formulas beyond $q=2$ requires carefully counting powers of $n$ to establish which terms survive in the $n\rightarrow\infty$ limit, and using the general expressions for $f_q$ and $g_q$ (Eqs.~\ref{eq:fqDefinition} and \ref{eq:gqDefinition}) to derive explicit forms of the leading-order terms. 
To tame this sum our derivation thus uses techniques that go beyond a simple generalization of \cite{farhi2019quantum}. 

\subsection{Evaluating Sums Over the Sketches $(n_{++},n_{+-},n_{-+},n_{--})$}

We see that evaluating both moments reduces to repeatedly evaluating terms of the form
\begin{align} 
    T^{ab}_\xi & :=\frac{1}{n^\xi}\sum_{\{n_*\}}\binom{n}{n_*} \exp\left(-\sum_{q'=1}^d \gamma^2 g_{q'}(n_*)\frac{\sigma_{q'}^2}{2n^{q'-1}}\right)(n_{+-}-n_{-+})^a(n_{++}-n_{--})^b\smashoperator[r]{\prod_{ss'\in\{\pm\}^2}} Q_{ss'}^{n_{ss'}}\label{eq:TabDefn}
\end{align}
with $\xi \geq a+b$ playing the role of $q-c$ in Eq.~\ref{eq:firstMomentForm}.  
For a fixed $n$, note that $g_{q'}(n_*)$ only depends on $n_{+-}+n_{-+}$ (Eq.~\ref{eq:gqDefinition}) and with this symmetry we define the single variable function $\bar{g}_{q'}(n_{+-}+n_{-+}) := g_{q'}(n_{++},n_{+-},n_{-+},n_{--})$ and use $\sum_t$ with $t=n_{+-}+n_{-+}$. 
The remaining two degrees of freedom are further reorganized  by summing over $n_{+-}$ using $\sum_{n_{+-}+n_{-+}=t}$, and over $n_{++}$ using $\sum_{n_{++}+n_{--}=n-t}$.
We thus obtain 
\begin{align}
\begin{split}
T^{ab}_\xi=\frac{1}{n^\xi}\sum_{t=0}^n\binom{n}{t}\exp\left(-\sum_{q'=1}^d \gamma^2 \bar g_{q'}(t)\frac{\sigma_{q'}^2}{2n^{q'-1}}\right)&\underbrace{\left[ \sum_{n_{+-}+n_{-+}=t}\binom{t}{n_{+-}}(n_{+-}-n_{-+})^a Q_{+-}^{n_{+-}}Q_{-+}^{n_{-+}}\right]}_{A^a_t}\\&\qquad\times\underbrace{\left[\sum_{n_{++}+n_{--}=n-t}\binom{n-t}{n_{++}} (n_{++}-n_{--})^b Q_{++}^{n_{++}}Q_{--}^{n_{--}}\right]}_{B^b_t},
\label{eq:ExpandedTerm}
\end{split}
\end{align}
where we have defined $A^a_t$ and $B^b_t$ to be analyzed next in the limit of large $n$. 

\subsection{Large $n$ Limit of $A^a_t$, $B^b_t$, and $T^{ab}_\xi$}
As we plan to take the limit $n\rightarrow\infty$, we will use big $O(\cdot)$ notation in $n$ to keep track of the significant terms in $A^a_t$ and $B^b_t$ for constant $a$, $b$, and $t$.
The summations for $A^a_t$ and $B^b_t$ can be expressed exactly via the following identity (which is a generalization of the identities given in \cite{farhi2019quantum} for the case of $a=1,2$ and which can be proven by induction on $a$):
\begin{align}
    \sum_{n_1+n_2=u}\binom{u}{n_1}(n_1-n_2)^aQ_1^{n_1}Q_2^{n_2}& =(x\partial_x-y\partial_y)^a(x+y)^u\left|_{\substack{x=Q_1\\y=Q_2}}\right.
\end{align}
For $A^a_t$, we thus have
\begin{align}
    A^a_t & = (x\partial_x-y\partial_y)^a(x+y)^t\left|_{\substack{x=Q_{+-}\\y=Q_{-+}}}\right.
\end{align}
Using the operator equality $\partial_x\cdot x = 1+x\partial_x$ we see that expanding $(x\partial_x-y\partial_y)^a$ gives a sum of differential operators $x^ky^\ell\partial_x^k\partial_y^\ell$ with $k+\ell\leq a$. 
Because $(Q_{+-}+Q_{-+})=0$ (Eq.~\ref{eq:QDefinition}) we see that only the operators $\partial_x^k\partial_y^\ell$ with $k+\ell=t$ give nonzero contributions to $A^a_t$. 
For $t>a$ we therefore have $A^a_t=0$.
In the particular case of $t=a$, the only terms in $(x\partial_x-y\partial_y)^a$ that give a nonzero result are $\sum_{k+\ell =a}\left(\begin{smallmatrix}a\\k\end{smallmatrix}\right) x^k(-y)^\ell\partial_x^k \partial_y^\ell $, hence
\begin{align}
    A_{a,a}= a!\sum_{k+\ell =a}\binom{a}{k} Q_{+-}^k(-Q_{-+})^\ell =a!(Q_{+-}-Q_{-+})^a=a!(-i)^a\sin^a (2\beta).
\end{align}
For $t<a$ it will be enough to observe that the order of $A^a_t$ is independent of $n$.
Summarizing, we have 
\begin{align}
    A^a_t =
    \begin{cases}
    0 & \text{if $t>a$,} \\
    a!(-i)^a\sin^a(2\beta) 
    & \text{if $t = a$,}\\
    O(1) & \text{if $t<a$}.
    \end{cases}
    \label{eq:At}
\end{align}

For $B^b_t$, we use
\begin{align}
    B^b_t & = (x\partial_x-y\partial_y)^b(x+y)^{n-t}\left|_{\substack{x=Q_{++}\\y=Q_{--}}}\right..
\end{align}
The highest-order terms in $n$ come from the terms $\sum_{k+\ell =b}\left(\begin{smallmatrix}b\\k\end{smallmatrix}\right) x^k(-y)^\ell\partial_x^k \partial_y^\ell$ in $(x\partial_x-y\partial_y)^b$ and thus
\begin{align}
\begin{split}
    B^b_t &= \frac{(n-t)!}{(n-t-b)!}\sum_{k+\ell =b}\binom{b}{k} Q_{++}^k(-Q_{--})^\ell(Q_{++}+Q_{--})^{n-t-b}+ O(n^{b-1})\\
    &= n^b(Q_{++}-Q_{--})^{b}(Q_{++}+Q_{--})^{n-t-b}+ O(n^{b-1})\\
    &= n^b\cos^b(2\beta)+ O(n^{b-1}).\label{eq:Bt}
\end{split}
\end{align}

Plugging our Eqs.~\ref{eq:At} and \ref{eq:Bt} for $A^a_t$ and $B^b_t$ into Eq.~\ref{eq:ExpandedTerm} for $T_\xi^{ab}$ hence gives
\begin{align}
\begin{split}
T^{ab}_\xi&=\frac{1}{n^\xi} \binom{n}{a}\exp\left(-\sum_{q'=1}^d \gamma^2 \bar{g}_{q'}(a)\frac{\sigma_{q'}^2}{2n^{q'-1}}\right)\cdot 
    a!(-i)^a\sin^a(2\beta) 
\cdot \left[n^b\cos^b(2\beta)+ O(n^{b-1})\right] \\
& \quad + \frac{1}{n^\xi} \sum_{t=0}^{a-1}\binom{n}{t}\exp\left(-\sum_{q'=1}^d \gamma^2 \bar{g}_{q'}(t)\frac{\sigma_{q'}^2}{2n^{q'-1}}\right)\cdot O(1)\cdot \left[n^b\cos^b(2\beta)+ O(n^{b-1})\right].
\end{split}
\end{align}
As $\binom{n}{a} = n^a/a!+O(n^{a-1})$ and similarly  $\binom{n}{t} = O(n^t)$ this simplifies to
\begin{align}
T^{ab}_\xi&= \frac{n^{a+b-\xi}}{a!}\exp\left(-\sum_{q'=1}^d \gamma^2 \bar{g}_{q'}(a)\frac{\sigma_{q'}^2}{2n^{q'-1}}\right)\cdot a!(-i)^a\sin^a(2\beta) \cdot \cos^b(2\beta)+O\left({n^{a+b-\xi-1}}\right).
\end{align}
After reminding ourselves that $a+b \leq \xi$ we see that, in the $n\rightarrow\infty$  limit, $T_\xi^{ab}$ further condenses to
\begin{align}
T^{ab}_\xi & \rightarrow
\begin{cases}
(-i)^a\exp\left(-\sum_{q'=1}^d \gamma^2 \sigma_{q'}^2\lim_{n\rightarrow\infty}\left[\frac{\bar{g}_{q'}(a)}{2n^{q'-1}}\right]\right)\sin^a(2\beta)\cos^b(2\beta) & \text{if~}a+b=\xi\\ 0 & \text{if~}a+b<\xi.
\end{cases}
\label{eq:Tab} 
\end{align}

As $\bar{g}_{q'}(a)$ has an implicit $n$ dependency we have to determine its relevant terms as $n\rightarrow\infty$. 
Eq.~\ref{eq:Tab} shows that the only relevant terms in $\bar{g}_{q'}$ are those of degree at least $(q'-1)$ in $n$. 
From the definition of $g_q$ (Eq.~\ref{eq:gqDefinition}), elementary algebra gives 
\begin{align}
    \bar{g}_{q'}(n_{+-}+n_{-+}) & =\frac{4(n_{+-}+n_{-+})n^{q'-1}}{(q'-1)!}+O(n^{q'-2}).
\end{align}
Therefore, we have the large $n$ limit 
\begin{align}
T^{ab}_{\xi} & \rightarrow
\begin{cases}
(-i)^a\exp\left(-2a\gamma^2\sum_{q'=1}^d  \frac{\sigma_{q'}^2}{(q'-1)!}\right)\sin^a(2\beta)\cos^b(2\beta) & \text{if~}a+b=\xi\\ 0 & \text{if~}a+b<\xi.
\end{cases}
\label{eq:Tab-final} 
\end{align}

\subsection{Large $n$ properties of $f_q$}
To complete the evaluation of the moments, we to determine the large $n$ dependency of $f_q$, its $f_q^{abc}$ coefficients (Eq.~\ref{eq:fqExpansion}) and their role in the first and second moment expressions of Eqs.~\ref{eq:firstMomentForm} and \ref{eq:secondMomentForm}. 
We see from Eq.~\ref{eq:Tab-final} that the only coefficients  that matter  in the $n\rightarrow\infty$ limit are those with $a+b=\xi$, i.e.\ those with $a+b+c=q$. 
We can evaluate these $f_q^{abc}$ from the explicit formula for $f_q$ (Eq.~\ref{eq:fqDefinition}) by keeping only terms of total degree $q$ in the terms $(n_{+-}-n_{-+})$, $(n_{++}-n_{--})$, and $n$. 
This gives
\begin{align}
    f_q(n_{++}, n_{+-}, n_{-+}, n_{--}) & =\sum_{\substack{a\text{ odd}\\a\leq q}}\frac{2}{a!(q-a)!}(n_{+-}-n_{-+})^a(n_{++}-n_{--})^{q-a} + (\text{lower degree terms})
    \label{eq:FqProps}
\end{align}
Note that there are no terms $(n_{+-}-n_{-+})^a(n_{++}-n_{--})^bn^c$ with $c\neq 0$ and $(a+b+c)=q$ in $f_q$. 
Comparing Eq.~\ref{eq:FqProps} with the definition of $f_q^{abc}$ in Eq.~\ref{eq:fqExpansion}, we see for $n\rightarrow\infty$ after some algebra that 
\begin{align}
    f_q^{abc}& =
    \begin{cases}\frac{2}{a!b!} & \text{if $a+b=q$, $c=0$, and $a$ odd}\\
    0 & \text{if $a+b+c=q$ with $a$ even or $c>0$.}
    \end{cases}\label{eq:fqabc-final}
\end{align}
Having established these properties of $T^{ab}_{q-c}$ and $f_q$, we now have sufficient information to evaluate our moments.

\subsection{Evaluating the First Moment}

To evaluate the first moment, we simplify Eq.~\ref{eq:firstMomentForm} using the definition of $T_\xi^{ab}$ (Eq.~\ref{eq:TabDefn}) and the results of Eqs.~\ref{eq:Tab-final} and \ref{eq:fqabc-final} to get
\begin{align}
\begin{split}
    \E_{J\sim N}\left[\braket{H/n}\right] 
    &= i\gamma\sum_{q=1}^d\sigma_q^2\sum_{a+b+c\leq q}f_q^{abc}T_{q-c}^{ab}\\
    & \rightarrow i\gamma\sum_{q=1}^d\sigma_q^2\sum_{\substack{a\text{ odd}\\a\leq q}}\frac{2}{a!(q-a)!}(-i)^a \exp\left(-2\gamma^2a\sum_{q'=1}^d  \frac{\sigma_{q'}^2}{(q'-1)!}\right)\sin^a(2\beta)\cos^{q-a}(2\beta)\\
    & = 2\gamma\sum_{q=1}^d\sigma_q^2\sum_{\substack{a\text{ odd}\\a\leq q}}\frac{(-1)^{\frac{a-1}{2}}}{a!(q-a)!}\exp\left(-2\gamma^2a\sum_{q'=1}^d  \frac{\sigma_{q'}^2}{(q'-1)!}\right)\sin^a(2\beta)\cos^{q-a}(2\beta),
\end{split}
\end{align}
which is precisely what we claimed in Eq.~\ref{eq:firstMomentResult}.

\subsection{Evaluating the Second Moment}
To evaluate the second moment, we simplify Eq.~\ref{eq:secondMomentForm}  using the definition of $T_\xi^{ab}$ (Eq.~\ref{eq:TabDefn}) and the results of Eqs.~\ref{eq:Tab-final} and \ref{eq:fqabc-final} to get
\begin{align}
\begin{split}
    \E_{J\sim N}\left[\braket{(H/n)^2}\right] &=
    \sum_{q=1}^d\binom{n}{q}\frac{\sigma_{q}^2}{n^{q+1}}-\gamma^2\sum_{q,q'=1}^d\sigma_q^2\sigma_{q'}^2\sum_{\substack{a+b+c\leq q\\a'+b'+c'\leq q'}}f^{abc}_q f^{a'b'c'}_{q'}T^{(a+a')(b+b')}_{q-c+q'-c'}\\
    &\rightarrow -\gamma^2\sum_{q,q'=1}^d\sigma_q^2\sigma_{q'}^2\sum_{\substack{a,a'\text{ odd}\\a\leq q\\a'\leq q'}}\frac{2}{a!(q-a)!}\frac{2}{a'!(q'-a')!}(-i)^{a+a'}\\&\qquad\qquad\qquad\qquad\qquad\qquad\times\exp\left(-2\gamma^2(a+a')\sum_{q''=1}^d \frac{\sigma_{q''}^2}{(q''-1)!}\right)\sin^{a+a'}(2\beta)\cos^{q-a+q'-a'}(2\beta)\\
    &= \left[i\gamma\sum_{q=1}^d\sigma_q^2\sum_{\substack{a\text{ odd}\\a\leq q}}\frac{2}{a!(q-a)!}(-i)^{a}\exp\left(-2\gamma^2a\sum_{q''=1}^d \frac{\sigma_{q''}^2}{(q''-1)!}\right)\sin^{a}(2\beta)\cos^{q-a}(2\beta)\right]^2\\
    &=\lim_{n\rightarrow\infty} \E_{J\sim N}\left[\braket{H/n}\right]^2,
\end{split}
\end{align}
which is precisely what we claimed in Eq.~\ref{eq:SecondMomentResult}.

\subsection{Evaluating Higher Moments}

The proof technique used above also applies to higher moments. When computing the $m$th moment by taking derivatives of the moment-generating function according to Eq.~\ref{eq:momentDefn}, the only terms that survive in the limit $n\rightarrow\infty$ are the terms where all derivatives hit the $2\gamma\lambda f_q /n$ term in the exponential, so that the expression for the moment becomes
\begin{align}
\E_{J\sim N}\left[\braket{(H/n)^m}\right]&= (i\gamma)^m\sum_{q_1,\dots,q_m=1}^d\sigma_{q_1}^2\cdots \sigma_{q_m}^2\sum_{\substack{a_1+b_1+c_1\leq q_1\\\dots\\a_m+b_m+c_m\leq q_m}}f^{a_1b_1c_1}_{q_1}\cdots f^{a_mb_mc_m}_{q_m}T^{(a_1+\cdots+a_m)(b_1+\cdots +b_m)}_{(q_1-c_1)+\cdots +(q_m-c_m)}+O\left(\frac{1}{n}\right).
\end{align}
Noting from Eq.~\ref{eq:Tab-final}  that
\begin{equation}
    T^{(a_1+\cdots+a_m)(b_1+\cdots +b_m)}_{(q_1-c_1)+\cdots +(q_m-c_m)} = T^{a_1b_1}_{q_1-c_1}\cdots T^{a_mb_m}_{q_m-c_m}+O\left(\frac{1}{n}\right),
\end{equation}
we can write the $m$th moment as
\begin{align}
\begin{split}
\E_{J\sim N}\left[\braket{(H/n)^m}\right]&= (i\gamma)^m\sum_{q_1,\dots,q_m=1}^d\sigma_{q_1}^2\cdots \sigma_{q_m}^2\sum_{\substack{a_1+b_1+c_1\leq q_1\\\cdots\\a_m+b_m+c_m\leq q_m}}f^{a_1b_1c_1}_{q_1}\cdots f^{a_mb_mc_m}_{q_m}T^{a_1b_1}_{q_1-c_1}\dots T^{a_mb_m}_{q_m-c_m}+O\left(\frac{1}{n}\right)\\
&= \left[i\gamma\sum_{q=1}^d\sigma_{q}^2\sum_{a+b+c\leq q}f^{abc}_{q}T^{a}_{q-c}\right]^m +O\left(\frac{1}{n}\right)\\
&=\E_{J\sim N}\left[\braket{H/n}\right]^m+O\left(\frac{1}{n}\right).
\end{split}
\end{align}
Thus, we find that for all $m$,
\begin{align}
    \lim_{n\rightarrow\infty}\E_{J\sim N}\left[\braket{(H/n)^m}\right] & =    \lim_{n\rightarrow\infty}\E_{J\sim N}\left[\braket{H/n}\right]^m .
\end{align}

\subsection{Alternative expression for $\mathbb{E}\left[\braket{H/n}\right]$}
Here, we present an alternative expression for $\mathbb{E}\left[\braket{H/n}\right]$ that makes contact with the notation used in \cite{alaoui2020algorithmic} and elsewhere (see Section~\ref{sec:PriorResults}). 
The current paper's central result (Eq.~\ref{eq:firstMomentResult}) reads:
\begin{align}
    \lim_{n\rightarrow\infty}\E_{J\sim N}\left[\braket{H/n}\right]&=2\gamma\sum_{q=1}^d\sigma_q^2\sum_{\substack{a\text{ odd}\\a\leq q}}\frac{(-1)^{\frac{a-1}{2}}}{a!(q-a)!}\exp\left(-2\gamma^2a\sum_{q'=1}^d  \frac{\sigma_{q'}^2}{(q'-1)!}\right)\sin^a(2\beta)\cos^{q-a}(2\beta).
\end{align}
If we use the substitution from Section~\ref{sec:PriorResults}, $\sigma^2_q = c^2_q {q!}$, this becomes
\begin{align}
\begin{split}
    \lim_{n\rightarrow\infty}\E_{J\sim N}\left[\braket{H/n}\right]&=2\gamma\sum_{q=1}^d c_q^2 q!\sum_{\substack{a\text{ odd}\\a\leq q}}\frac{(-1)^{\frac{a-1}{2}}}{a!(q-a)!}\exp\left(-2\gamma^2a\sum_{q'=1}^d  q'c_{q'}^2\right)\sin^a(2\beta)\cos^{q-a}(2\beta)\\
    & = 
    2i\gamma\sum_{q=1}^d c^2_q \sum_{\substack{a\text{ odd}\\0\leq a\leq q}}\binom{q}{a}\left(-i \sin(2\beta) \exp\left({-2\gamma^2\sum_{q'=1}^d  q' c^2_{q'}}\right)\right)^a\big(\cos(2\beta)\big)^{q-a}.
\end{split}
\end{align}
Using the identity
\begin{align}
2i\sum_{a\text{ odd}}\binom{q}{a}(-iA)^a B^{q-a} & = -i ((iA+B)^q-(-iA+B)^q)  = 2\Im\left[(iA+B)^q\right], 
\end{align}
we can rewrite this as
\begin{align}
    \lim_{n\rightarrow\infty}\E_{J\sim N}\left[\braket{H/n}\right]    & = 
    2\gamma\sum_{q=1}^d c^2_q 
    \Im\left[
    \left(
    i \sin(2\beta) \exp\left({-2\gamma^2\sum_{q'=1}^d  q' c^2_{q'}}\right)
    +
    \cos(2\beta)
    \right)^q
    \right].
\end{align}
Finally, using the definition $\xi(x) := \sum_q c_q^2 x^q$ and hence $\xi'(1) = \sum_q q c_q^2$, we get 
\begin{align}
\begin{split}
    \lim_{n\rightarrow\infty}\E_{J\sim N}\left[\braket{H/n}\right]    & = 
    2\gamma\sum_{q=1}^d c^2_q 
    \Im\left[
    \Big(
       \cos(2\beta) + 
    i \sin(2\beta) \exp\left({-2\gamma^2 \xi'(1)}\right)
    \Big)^q
    \right]\\
    & = 
    2\gamma\Im\left[\xi 
    \Big(
       \cos(2\beta) + 
    i \sin(2\beta) \exp\left({-2\gamma^2 \xi'(1)}\right)
    \Big)\right],
\end{split}
\end{align}
which is precisely what we claimed in Eq.~\ref{eq:firstMomentResultMontanariForm}. For a pure $d$-spin model with only $c_d\neq 0$ the above further simplifies to
\begin{align}
    \lim_{n\rightarrow\infty}\E_{J\sim N}\left[\braket{H/n}\right]    & = 
    2\gamma c_d^2\cdot
    \Im\left[
    \Big(
       \cos(2\beta) + 
    i \sin(2\beta) \exp\left({-2dc_d^2 \gamma^2}\right)
    \Big)^d
    \right].
\end{align}

\section{Discussion and Conclusion}

In this work, we have derived explicit formulas to quantify the performance of $p=1$ \QAOA on mixed-spin models in the $n\rightarrow\infty$ limit. 
We demonstrated both concentration of measure and instance independence for arbitrary mixed-spin models, which imply that the expectation value of the energy per spin is independent of the specific model specification and that measurements of the \QAOA state are guaranteed to give energies close to the expectation value. 
Our explicit formula for the expectation value of the energy for arbitrary mixed-spin models allows us to find the optimal angles on a classical computer.

There are two obvious open questions raised by this work. 
First, the approach of this paper can probably be combined with the methods of \cite{farhi2019quantum} to generalize our work to depth $p>1$ \QAOA. 
Such a result would allow one to prove instance independence and concentration of measure, and derive a computer algorithm to generate formulas for the expectation value per spin, at arbitrary depth $p$. 
This is a particularly interesting route of research, since it is known that in the cubic case with $\sigma_q\propto\delta_{q,3}$, Montanari's classical algorithm does not approach the optimal solution \cite{alaoui2020algorithmic}, so that at sufficient depth $p$ the \QAOA has a chance of outperforming the best known classical algorithm. 
Higher-spin models with $q>3$ may even provide a more direct route towards finding a setting where \QAOA at depth $p=1$ outperforms classical optimization such as Montanari's algorithm \cite{alaoui2020optimization,alaoui2020algorithmic}. 
While the generalization to higher $p$ is likely possible, it is a nontrivial extension of this paper, and we leave it for future work.

Second, it remains an open question what to what extent results on the random models can be used to find optimal angles for realistic binary optimization problems. 
One hypothetical approach to finding \QAOA angles for a single instance of an $n$-spin optimization problem would be the procedure:
\begin{enumerate}
    \item For all $q=1,\dots,d$, compute the standard deviations $\sigma_q$ of the $q$-spin couplings in the problem.
    \item Use Eq.~\ref{eq:firstMomentResult} for the mixed-spin model as an estimate for the expectation value of the energy per spin.
    \item Run the \QAOA at the optimal angles for the corresponding mixed-spin model (or use these angles as starting points for a numerical optimization of the angles)
\end{enumerate}
This procedure assumes that the means of the couplings are zero: $\mu_q=0$ for all $q=1,\dots,d$.  
If this method can be shown to work, it would greatly simplify finding \QAOA angles for general optimization problems.

\bibliographystyle{unsrtnat}
\bibliography{references}

\appendix

\end{document}